\documentclass[12pt,doublespace]{article}
\begin{document}
\begin{center}
DITHERED TIME DELAYS AND CHAOS SYNCHRONIZATION IN LASERS WITH MULTIPLE FEEDBACKS \\
E.M.Shahverdiev $^{1,2}$ and K.A.Shore $^{1}$\\
$^{1}$School of Electronic Engineering,Bangor University, Dean St.,Bangor, LL57 1UT, Wales, UK\\
$^{2}$Institute of Physics, H.Javid Avenue,33, Baku, AZ1143, Azerbaijan\\
ABSTRACT
\end{center}
By studying laser systems with multiple time delays, we demonstrate that the signatures of time delays in the autocorrelation coefficient and  the mutual information of the laser output can be erased for systems with variable time delays. This property makes such laser systems highly suitable for secure chaos-based communication systems. We also present the first report on chaos synchronization in both unidirectionally and bidirectionally coupled variable multiple time delay laser diodes with electro-optical feedbacks.\\ 
PACS number(s):42.55.Px, 42.65.Sf, 05.45.Xt, 42.60.Mi,05.45.Gg, 05.45.Vx\\
{\it Introduction.}-In recent years due to its fundamental and applied interests chaos synchronization has been the subject of extensive studies [1-2]. From the application point of view, chaos based communication systems  can offer improved privacy and security in data transmission, especially after the recent field demonstration using 
a metropolitan fibre network [3]. In optical chaos based communications, the chaotic waveform is generated by using semiconductor lasers with all-optical or electro-optical feedback(s). Mathematically such time delayed systems are described by delay differential equations (DDE)[4]. Recent studies have revealed that DDE-chaos based communication systems [1-2] could be more secure than one based on chaos in ODE (ordinary differential equations) [5], as DDEs  usually have a very high dimensionality and many positive Lyapunov exponents. However, the authors of [6], were successful in extracting messages masked by a chaotic signal of a time delay system. In [6] the authors used the average fitting error as a function of the embedding delay, which had a pronounced minimum at the time delay. This allowed them to correctly identify the time delay of the system and to extract the message successfully using a simple local reconstruction of the time delay system. The delay time can also be exposed by several other measures, e.g. autocorrelation coefficient [7] and mutual information [8].\\
\hspace*{0.5cm} Quite recently it was suggested that multiple time delay systems can offer a higher complexity of dynamics than is achievable in more conventional single delay time systems [9]. The role of additional time delays in achieving a homogeneous steady state in coupled chaotic maps was investigated in recent work [10] with the stabilizing role of the additional time delays being emphasized. The stabilizing role of additional time delays 
in external cavity laser diodes with multiple feedbacks is highlighted in [11]. The high complexity of such multiple time delay systems can provide a new architecture for enhancing message security in chaos based encryption systems. However using the same approach as in [6], it was shown that for the  multiple feedback systems, time delays can also be successfully recovered and reconstruction of the dynamics of the systems was still possible [12]. A viable recent proposal to enhance the security of the chaos-based communication schemes is the use of time delayed systems in which the delay time is modulated in time [13].\\
\indent In this paper we study semiconductor lasers with modulated multiple time delay electrooptical feedbacks. First we show that contrary to the case of constant time delays, with variable time delay systems the delays 
cannot be recovered by investigating the autocorrelation coefficient or the mutual information of the laser output. Thus variable time delay systems can offer an increase in the security of chaos based communication systems. For message decoding in such schemes one has to be able to synchronize the transmitter and receiver lasers. In the light of this, in the paper we also present the first report on chaos synchronization in both unidirectionally and bidirectionally coupled variable multiple time delayed semiconductor lasers with electro-optical feedbacks.\\ 
{\it Wavelength chaos model.}-The laser system considered in this paper is an electrically tunable Distributed Bragg Reflector (DBR) laser diode with a feedback loops, whose property is to exhibit a nonlinearity in wavelength. This system was proposed in [14] as a chaotic wavelength signal generator for chaos based secure communication.\\
The output wavelength of the chaotic oscillator is described by the following
dynamical equation: $T\frac{\lambda(t)}{dt}=-\lambda (t) + \beta_{\lambda}\sin^{2}(\frac{D\pi}{\Lambda^{2}_{0}}\lambda (t-\tau) -\Phi_{0}),$
where $\lambda$ is the wavelength deviation from the center wavelength $\Lambda_{0}$;$D$ is the optical path difference
of the birefringent plate that provides the nonlinearity;$\Phi_{0}$ is the feedback phase;$\tau$-the feedback loop
delay time;$T$ is the time response in the feedback loop;$\beta_{\lambda}$ is the feedback strength. With $x=(\pi D\lambda)\Lambda^{-2}_{0}$ and $m=(\pi D\beta_{\lambda})\Lambda^{-2}_{0}$ we rewrite it in the following normalized  form:$\frac{dx(t)}{dt}=-\alpha x(t) + m \sin^{2}(x_{\tau} - \Phi_{0}),$ where $\alpha$ is the relaxation coefficient and $x_{\tau} \equiv x(t-\tau).$ In this paper we will consider the variable time delay case $\tau= \tau (t).$ In the following we consider chaos synchronization between both unidirectionally and bidirectionally coupled identical laser diodes with double feedback. In figure 1  a schematic diagram of the experimental set-up for synchronization is given.\\ 
The principal elements of the chaotic oscillator are:\\
(1) a tunable laser source (DBR) whose wavelength can be tuned continuously, i.e. 
by a DBR-section injection current $I;$\\
(2) a wavelength nonlinear element formed by a birefringent plate (BP) set between two crossed polarizers, inducing $\sin^{2}$-nonlinearity;\\
(3) a photodetector (PD) providing a linear conversion of the optical power into a photocurrent  
with a conversion factor $G,$ which can be adjusted electronically;\\
(4) a delay line (DL) which introduces a time delay much longer than the response time of the feedback loop, in order 
to obtain the chaotic regime;\\
(5) a first-order low-pass filter (LPF), which determines the response time of the feedback loop.\\
Generally, an experimental realization of the variable time delays can be achieved by changing of the distance between the light source and photodetector, e.g. by using trombone structure, as for example in [15]. \\
{\it Numerical simulations.}-First we consider the case of unidirectionally coupled laser systems:
The master laser is described by the following equation
\begin{equation}
\frac{dx(t)}{dt}=-\alpha_{1} x(t) + m_{1} \sin^{2}(x_{\tau_{1}} -\Phi_{0}) + m_{2} \sin^{2}(x_{\tau_{2}} -\Phi_{0}),
\end{equation}
The dynamics of the slave laser is  governed by the following equation:
\begin{equation}
\frac{dy(t)}{dt}=-\alpha_{2} y(t) + m_{3} \sin^{2}(y_{\tau_{1}} -b\Phi_{0}) + m_{4} \sin^{2}(y_{\tau_{2}} -\Phi_{0}) + K \sin^{2}(x_{\tau_{3}} -\Phi_{0}),
\end{equation}
where $\tau_{1,2}=\tau_{01,02} + x_{1}(t) \tau_{a1,a2}\sin(\omega_{1,2}t)$ are the variable feedback loop delay times;$\tau_{3}=\tau_{03} + x_{1}(t)\tau_{a3}\sin(\omega_{3}t)$ is the variable time of flight between master laser $x$ and slave laser $y$;$\tau_{01,02,03}$ are the zero-frequency component,$\tau_{a1,a2,a3}$ are the amplitude,$\frac{\omega_{1,2,3}}{2\pi}$ are the frequency of the modulations; $x_{1}(t)$ is the output of laser (1) for constant time delays, i.e. $\tau_{1}=\tau_{01}, \tau_{2}=\tau_{02};$ $m_{1,2}$ and $m_{3,4}$ are the feedback strengths for the master and slave laser, respectively; $K$ is the coupling strength between the lasers. For  mutually coupled systems the term $K \sin^{2}(x_{\tau_{3}} -\Phi_{0})$ should be added to the right-hand side of Eq.(1). In the experimental scheme unidirectional coupling can be realized by inclusion of an optical isolator (OI), figure 1.\\
\hspace*{0.5cm}In the case of variable time delays establishing the existence and stability conditions for the synchronization is not as straightforward as for the constant time delays. Having in mind that for $\omega=0$ we obtain a case of constant time delays, then as an initial guess one can benefit from the existence conditions for the constant time delays case, see,e.g.[16]. It is our conjecture that high quality complete synchronization $x=y$ will be obtained if the parameters satisfy the conditions:$m_{1}=m_{3} + K, m_{2}=m_{4}$ if $\tau_{1}(t)=\tau_{3}(t)$ (i.e.the zero frequency components, amplitudes, and modulation frequencies are the same:$\tau_{01}=\tau_{03},\tau_{a1}=\tau_{a3},\omega_{1}=\omega_{3}$),or $m_{2}=m_{4} + K, m_{1}=m_{3}$ if $\tau_{2}(t)=\tau_{3}(t)$(i.e.$\tau_{02}=\tau_{03},\tau_{a2}=\tau_{a3},\omega_{2}=\omega_{3}$). As evidenced by the numerical simulations below, this conjecture is found to be well-based.\\
Before studying the synchronization between the laser systems with variable time delays we investigate the autocorrelation coefficient and mutual information  for the output of the master laser for both constant and variable time delays. The autocorrelation coefficient $C_{A}$ is a measure of how well a signal matches a time shifted version of itself and is a special case of the cross-correlation coefficient [17] $C(\Delta t)= <(x(t) - <x>)(y(t+\Delta t) - <y>)>(<(x(t) - <x>)^2><(y(t+ \Delta t) - <y>)^2>)^{-0.5}$ for $x=y:$ where $x$ and $y$ are the outputs of the interacting laser systems; the brackets$<.>$
represent the time average; $\Delta t$ is a time shift between laser outputs. This coefficient
indicates the quality of synchronization:C=0 implies no synchronization;$C={\pm 1}$ means perfect (inverse) synchronization.
The mutual information $J(\tau)$ between $x$ and $x_{\tau}$ is defined [8] by 
$J(\tau)=\sum_{x(t),x(t-\tau)} p(x(t),x(t-\tau))\log_{2}(p(x(t),x(t-\tau))(p(x(t))p(x(t-\tau)))^{-1},$ where $p(x(t),x(t-\tau))$ is the point probability and $p(x(t))$ and $p(x(t-\tau))$ are the marginal probability densities. The mutual information $J$ measures the information shared by two variables $x(t)$ and $x(t-\tau),$ namely it measures how  knowledge of one of these variables reduces the uncertainty about the other.\\
{\it Constant and variable time delay systems}\\
Figures 2(a) and 2(b) demonstrate the autocorrelation coefficient and the mutual information for the output of laser $x$ for constant time delays,respectively, i.e. for $\omega_{1}=\omega_{2}=0,\alpha_{1}=\alpha_{2}=4,\Phi_{0}=\pi/4, m_{1}=12, m_{2}=15, \tau_{01}=3,\tau_{02}=5.$ It is clearly seen that time delays can be easily recovered from both the autocorrelation coefficient and mutual information, as they exhibit extrema at time delays or their multiples and combinations. It is noted that the identification of delays in the case of two delayed feedbacks was reported in [18].\\
\hspace*{0.5cm}Next let us consider the variable time delays scenario. In investigating the behavior of the autocorrelation coefficient and the mutual information we have experimented with different types of variable time delays, among them:(a)sinusoidal modulations:$\tau_{1,2}=\tau_{01,02} + \tau_{a1,a2}\sin(\omega_{1,2}t);$ (b)chaotic modulations:$\tau_{1,2}=\tau_{01,02} + \tau_{a1,a2} x_{1}(t);$and (c)combined chaotic and sinusoidal modulations $\tau_{1,2}=\tau_{01,02} + x_{1}(t)\tau_{a1,a2}\sin(\omega_{1,2}t).$ Extensive numerical simulations have established that erasure of the signatures of time delays in the autocorrelation coefficient and the mutual information is 
best achieved for combined chaotic and sinusoidal modulations of $\tau(t).$
Figures 3(a) and 3(b) show the autocorrelation coefficient and the mutual information of the master laser output for sinusoidal modulations 
for $\tau_{1}(t)=3 + 0.03\sin(0.006t)$ and $\tau_{2}(t)=5 + 0.03\sin(0.006t)$ with other parameters as in figure 2. In this connection we emphasize that the retrieval of the peroidic time delay from experimental time series by use of the mutual information and modified filling factor 
was presented in [19]. The autocorrelation coefficient and the mutual information of the laser output for chaotic time delays for $\tau_{1}(t)=3 + 0.03x_{1}(t)$ and $\tau_{2}(t)=5 + 0.03x_{1}(t)$ are portrayed in figure 4(a) and 4(b). Here $x_{1}(t)$ is the $x$ laser output for constant time delays for parameters as in figure 2. Finally figures 5(a) and 5(b) depict the autocorrelation coefficient and the mutual information of the laser output for combined sinusoidal and chaotic time delays, Eq.(3) for $\tau_{1}(t)=3 + 0.03x_{1}(t)\sin(0.006t)$ and $\tau_{2}(t)=5 + 0.03x_{1}(t)\sin(0.006t)$ with the rest of parameters as for figure 2.\\
Here we also would like to emphasize the following point. As noted in [20], with increasing the feedback rate it could become difficult or even impossible 
to recover the time delay from the time series. With this in mind we have conducted numerical simulations with both low and high level feedback rates.
The numerical simulations have demonstrated that with the right choice of the sampling rate of the time series one can still retrieve the time delays even for the case of increased feedback rates. We also note that the feature underlined in [20] is not generic. As demonstrated in [21] for the semiconductor laser with a single optical feedback low feedback rates are more suitable than the higher feedback to make the delay indentification difficult.\\
Thus, it is evident that combined chaotic and sinusoidal modulations of time delays is most successful in eliminating the signatures of the time delays in the autocorrelation coefficient and the mutual information of the laser output. In other words modulation of the delay times gives rise to the loss of their signature in the autocorrelation coefficient and the mutual information, and therefore can improve the security of chaos based communication systems.\\
{\it Synchronization properties}\\
As mentioned above, in chaos based communication schemes synchronization between the transmitter and receiver lasers are vital for message decoding. With this in mind we present here the first report of chaos synchronization between variable time-delay lasers. In the literature there are a few papers on chaos synchronization between chaotic systems with variable time delays [13,22]; however those papers dealt with synchronization in maps, or simple  systems, such as Lorenz, Rossler and Mackey-Glass. These paradigm chaos models in nonlinear dynamics are of limited practical interest in applications in fast communication schemes.\\
Figures 6(a) and 6(b) portray complete chaos synchronization $x=y$ between unidirectionally coupled lasers, Eqs.(1) and (2) for variable feedback time delays 
$\tau_{1}(t)=3 + 0.03\sin(0.006t), \tau_{2}(t)=5 + 0.03\sin(0.006t)$ and  variable coupling time delay $\tau_{1}(t)=3 + 0.03\sin(0.006t)$ with parameter values as $\alpha_{1}=\alpha_{2}=4, m_{1}=12, m_{2}=m_{4}=15, m_{3}=0.5,  K=11.5,\Phi=\pi/4.$ 
Figures 7(a) and 7(b) depicts chaos synchronization between mutually coupled lasers $x$ and $y$ with the same variable feedback and coupling time delays as in figure 6 for parameters $\alpha_{1}=\alpha_{2}=4, m_{1}=m_{3}=12, m_{2}=m_{4}=15, K=12.5,\Phi=\pi/4.$(We recall that in the case of mutually coupled lasers the  
the right-hand side of Eq.(1) is augmented by a term $K \sin^{2}(x_{\tau_{3}} -\Phi_{0}).$) The values of the cross-correlation coefficients for both unidirectionally and bidirectionally coupled systems testify to the high quality chaos synchronization, which is vital for information processing in chaos-based communication systems.\\
\indent In this connection we also study the effect of parameter mismatches on the synchronization quality. It is noted that sensitivity of the synchronization to mismatches of the parameters can lead to a high level of security due to the difficulty to replicate the receiver laser, i.e. sensitivity to parameter mismatches increases the security of encryption. However, the internal parameters of the interacting laser diodes unlikely to match exactly even if they are produced from the same wafer. Moreover, the operating parameters cannot be perfectly controlled. In other words, in practical cases, synchronization must therefore occur also for small parameter mismatches.  Most importantly, an investigation of the effect of parameter mismatches on synchronization quality will enable determination of the most sensitive synchronization parameters.In our numerical simulations we allow a 5$\%$ mismatch between parameters.\\
In figure 8 the dependence of the cross-correlation coefficient $C$ between the unidirectionally coupled lasers on the ratio  
$\frac{m_{2}}{m_{4}}$
of the feedback strength of the transmitter $m_{2}$ to the feedback strength of the receiver $m_{4}$($\diamondsuit$), on the ratio $\frac{\alpha_{1}}{\alpha_{2}}$
 of the relaxation coefficient of the transmitter laser $\alpha_{1}$ to the relaxation coefficient of the receiver laser $\alpha_{2}$($\nabla$), on the ratio $\frac{\omega_{1}}{\omega_{3}}$
 of the frequency of the 
feedback modulation of the tranmitter $\omega_{1}$ to the frequency of modulation of the injection term $\omega_{3}$($\triangle$), and on the ratio $\frac{\tau_{01}}{\tau_{03}}$
  of the fixed time delay of the transmitter laser $\tau_{01}$ to the fixed time delay of the injection term $\tau_{03}$($\star$) is presented.\\
It is emphasized that for these cases the synchronization quality is quite robust for small parameter mismatches (1$\%$).
These results also demonstrate high sensitivity of the synchronization quality to the parameter mismatches of the frequency of feedback modulations and feedback times. We have also found the similar trend for the mutually coupled laser systems 
even with more robustness to the parameter mismatches.\\
\indent Finally we dwell on the possibility of chaos control via variable time delays. The numerical results have demonstrated that there is an optimal frequency of feedback modulations to erase the signatures of time delays in the autocorrelation coefficient and the mutual information of the chaotic laser output. We have established that for frequencies  much higher than the optimal one the variable feedback(s) can be used to control the chaotic behaviour, i.e. to convert such a behaviour to the fixed state (figure 9).\\
\indent Before concluding we also emphasize that the laser model under study is of the Ikeda type and can show multi-stability. As it is demonstrated in [23] the relationship between the achievability of the synchronization and multi-stability in time delayed systems is a quite interesting subject. This relationship deserves more detailed study for the time delayed laser systems which is beyond the scope of this paper.\\
\hspace*{0.5cm}To summarize, by investigating autocorrelation coefficient and the mutual information we have established that in terms of security considerations, variable multiple time delay laser systems offer significant advantages for chaos-based communication schemes. We have also reported  on chaos synchronization in both unidirectionally and bidirectionally coupled variable multiple time delay laser diodes. The results of the paper provide the basis for the use of lasers diodes with multiple variable time delays in chaos-based secure high-speed communication systems.\\
{\it Acknowledgements.}-This research was supported by a Marie Curie Action within the $6^{th}$ European Community Framework Programme Contract N MIF1-CT-2006-039927 and Contract N MIF2-CT-2007-039927-980065.\\
\begin{center}
Figure captions
\end{center}
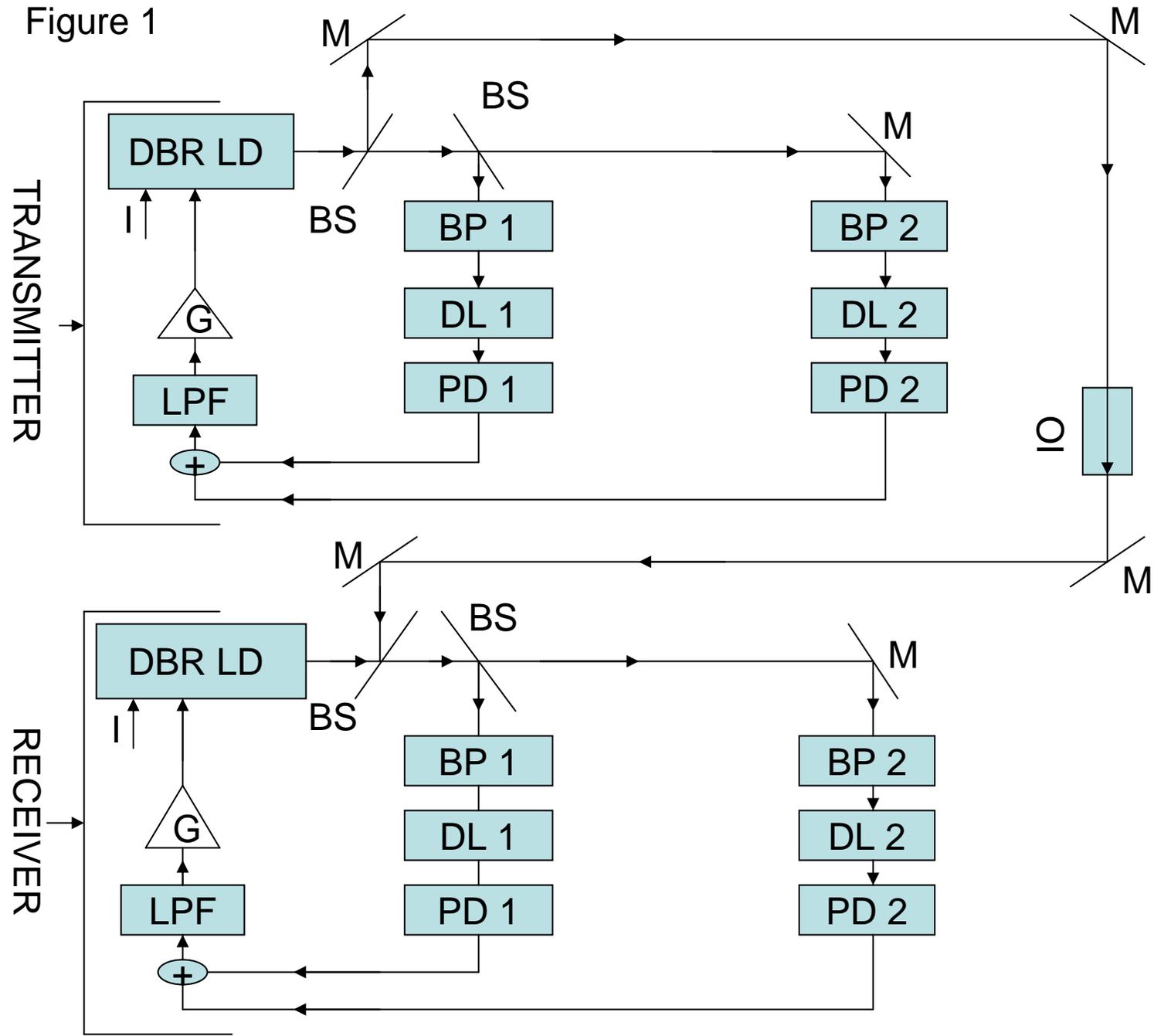
\noindent FIG.1. Schematic experimental set-up for wavelength chaos synchronization between the transmitter and receiver laser diodes with electooptical feedback: DBR LD1 and DBR LD2 are the Distributed Bragg Reflector transmitter and receiver laser diodes, respectively; BS, beamsplitter; M, mirror; BP, birefringent plate between crossed polarizers(not shown in the figure); PD, photodiode; DL, delay line; LPF, low-pass filter; G, optoelectronic gain; 
I, DBR-section injection current; OI, optical isolator to provide unidirectional coupling between laser diodes.\\
~\\
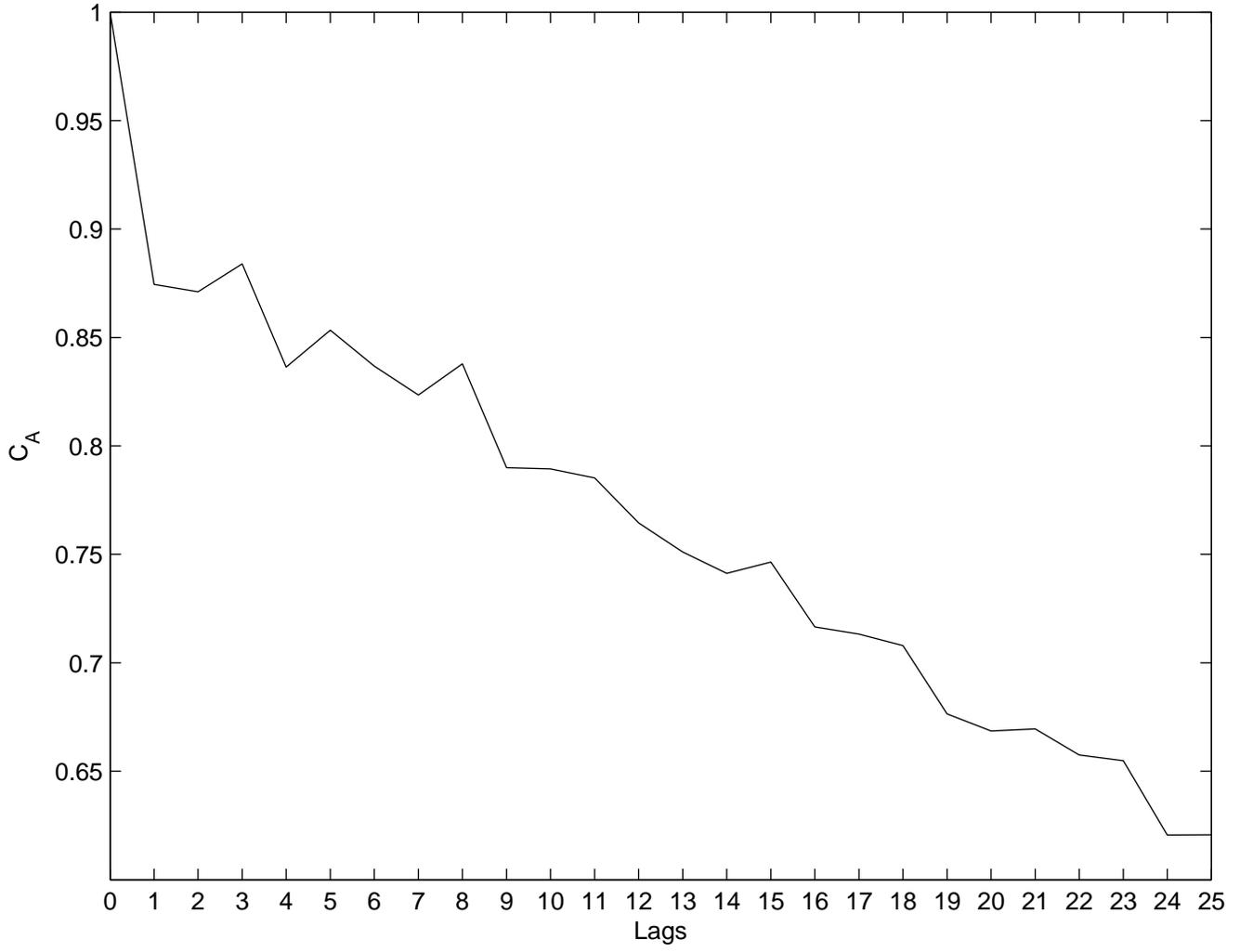
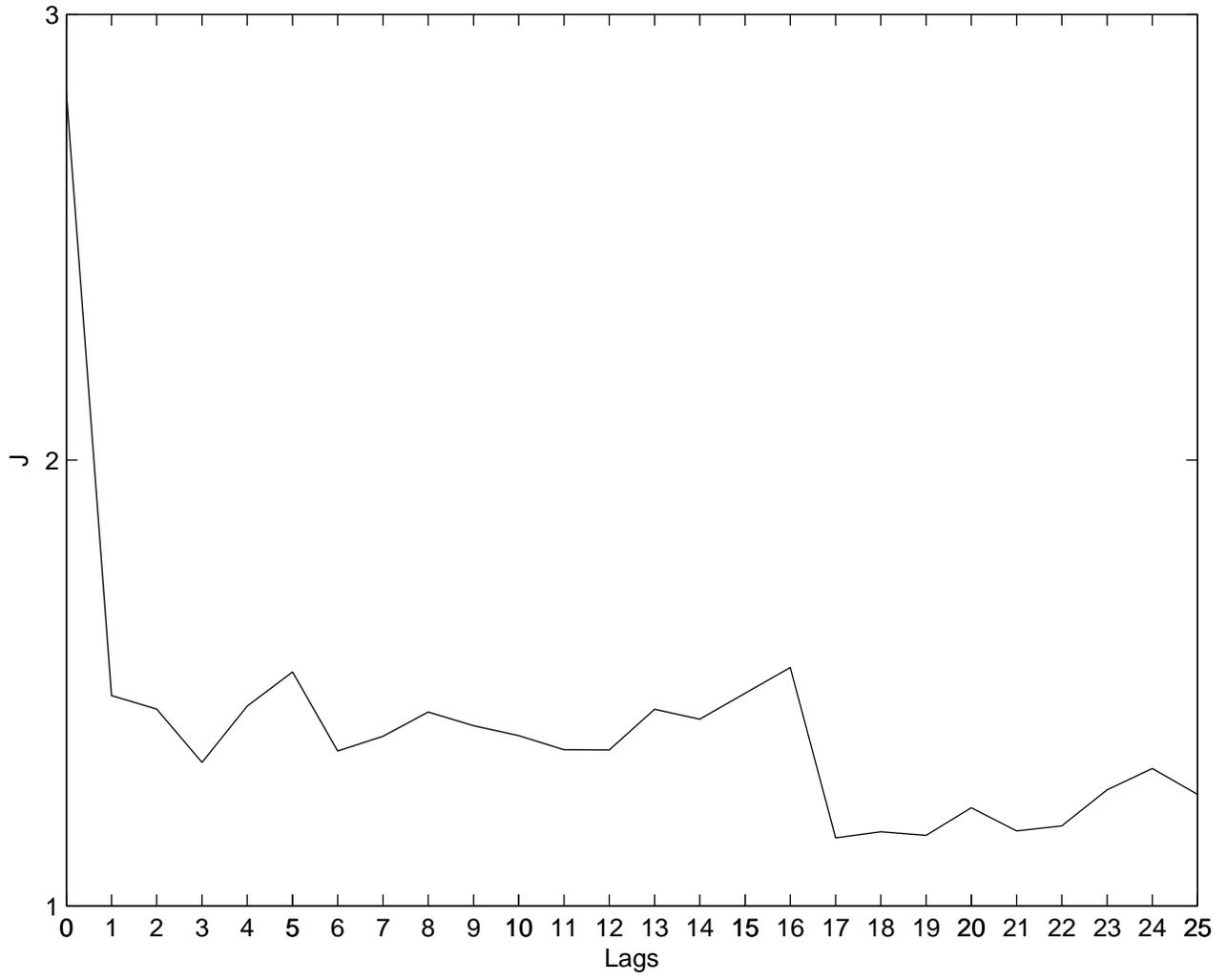
\noindent FIG. 2. The autocorrelation coefficient $C_{A}$ (2(a)) and mutual information $J$ (2(b)) of the laser output for constant time delays, Eq.(1) for $\alpha_{1}=\alpha_{2}=4, m_{1}=12, m_{2}=15,\tau_{01}=3,\tau_{02}=5,\Phi=\pi/4.$ Dimensionless units. \\
~\\
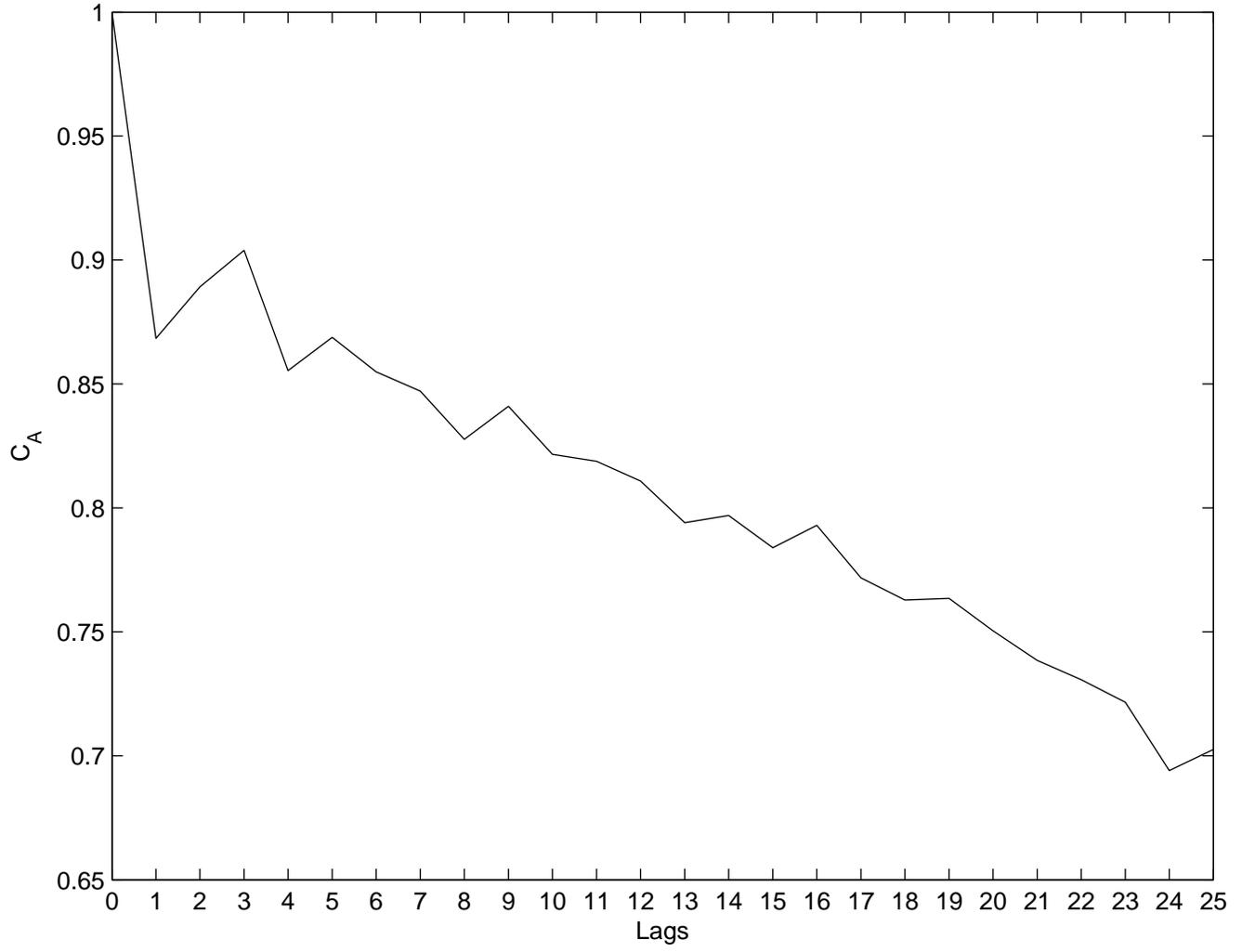
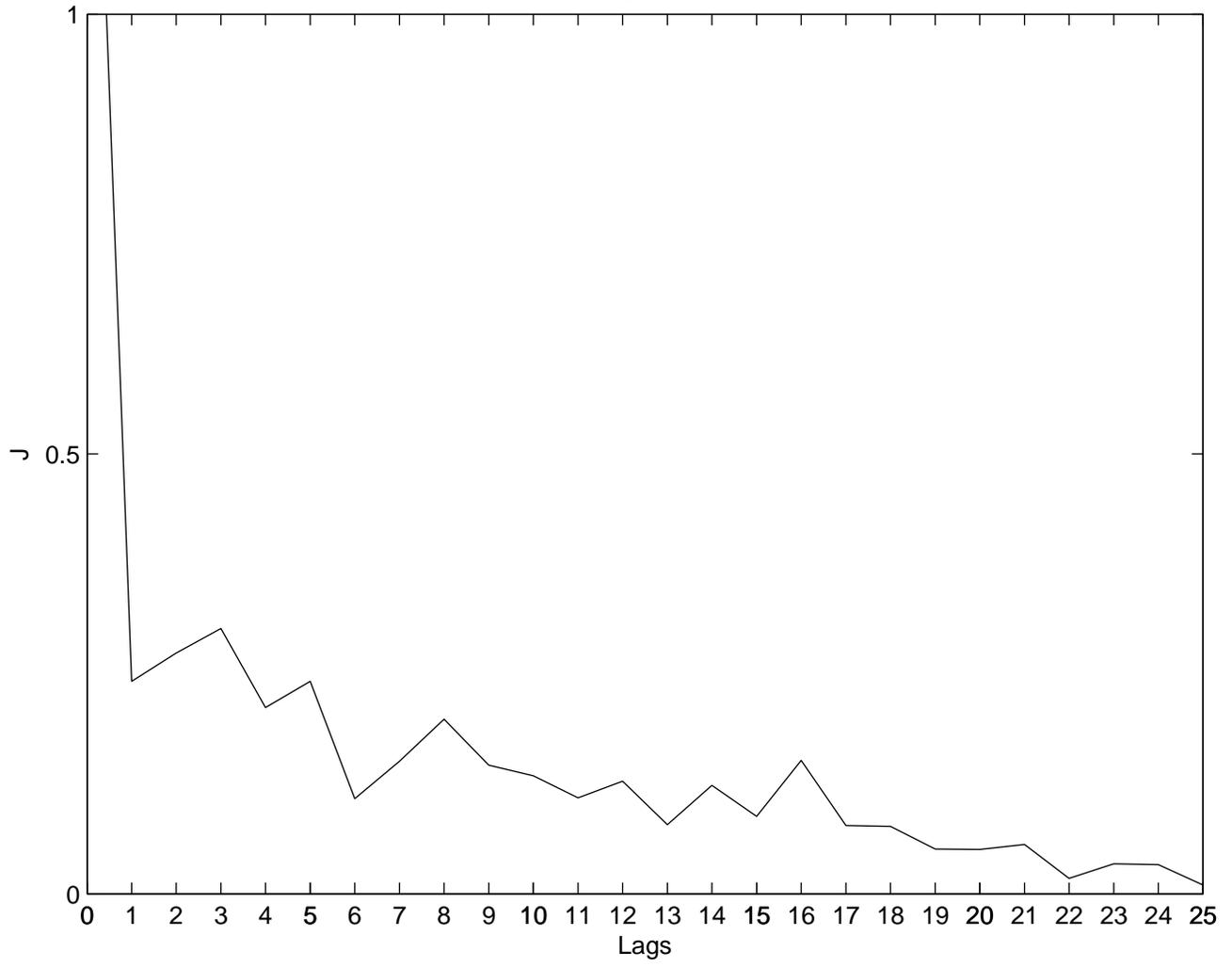
\noindent FIG.3. The autocorrelation coefficient $C_{A}$ (3(a)) and  mutual information $J$ (3(b)) of the laser output for sinusoidal modulations, Eq.(1) 
for $\tau_{1}(t)=3 + 0.03\sin(0.006t)$ and $\tau_{2}(t)=5 + 0.03\sin(0.006t).$ The other parameters are as in figure 2.Dimensionless units.\\
~\\
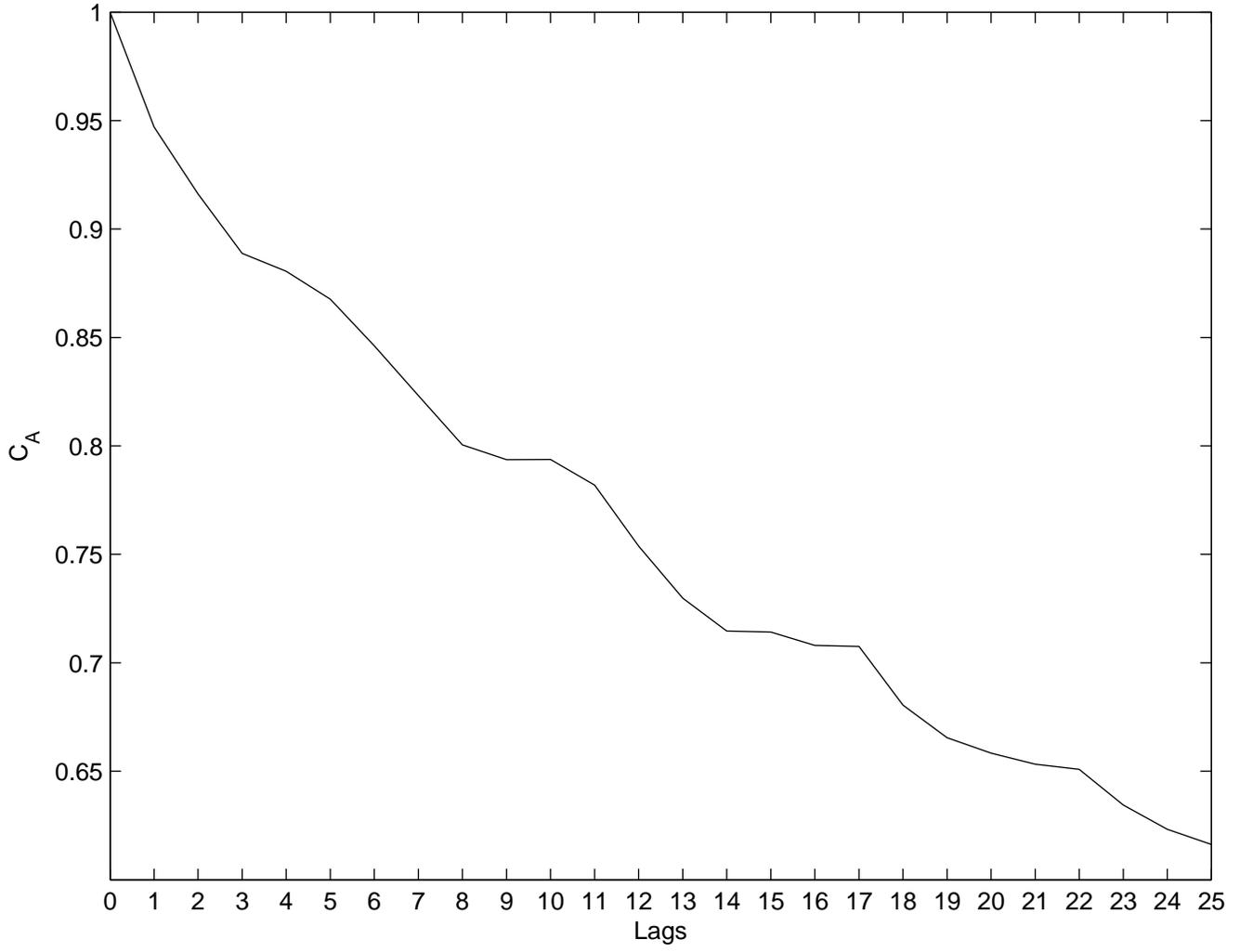
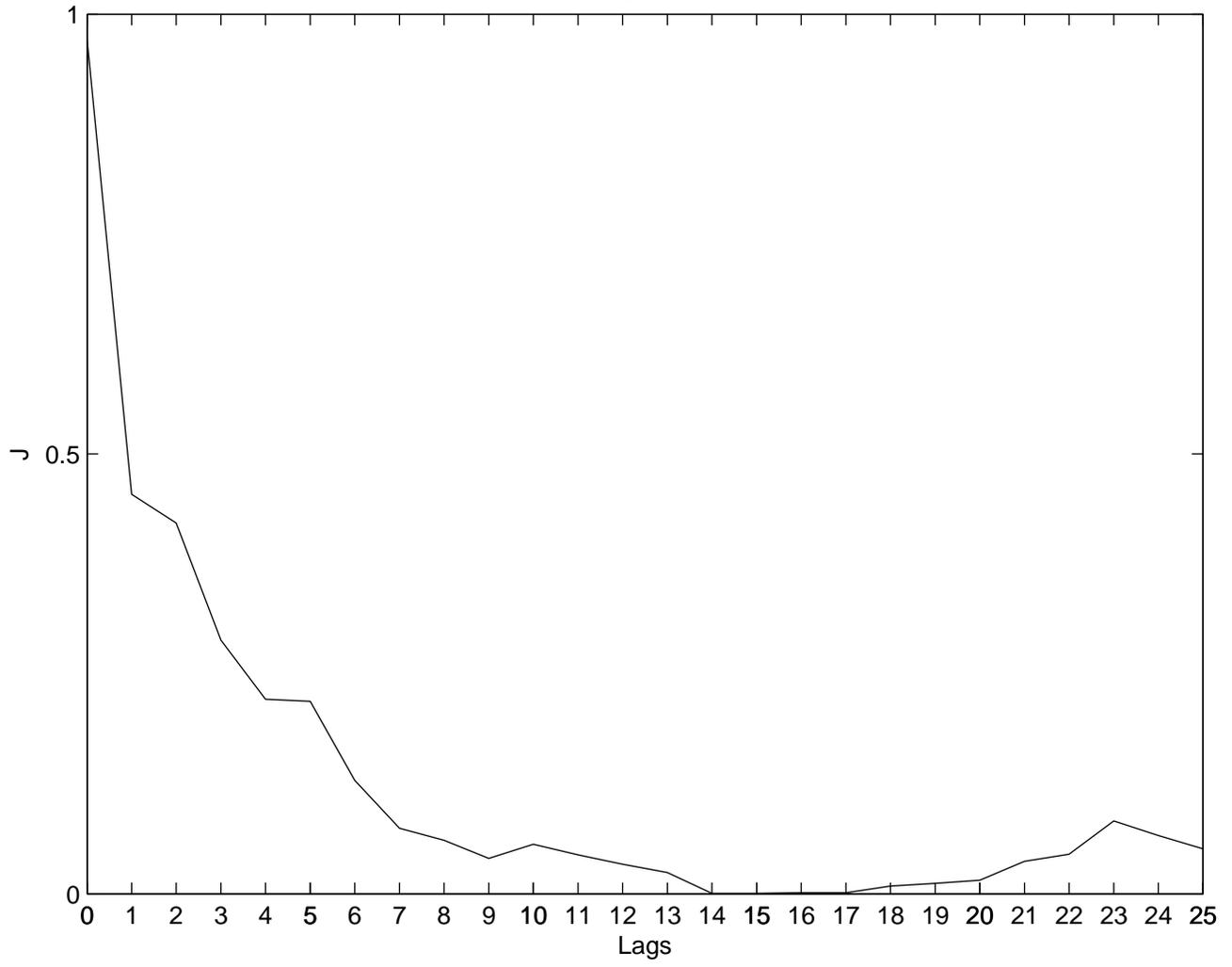
\noindent FIG.4. The autocorrelation coefficient $C_{A}$ (4(a)) and  mutual information $J$ (4(b)) of the laser output for chaotic time delays, Eq.(1) for 
$\tau_{1}(t)=3 + 0.03x_{1}(t)$ and $\tau_{2}(t)=5 + 0.03x_{1}(t).$ $x_{1}(t)$ is the laser output, Eq.(1) for constant time delays $\tau_{01}=3,\tau_{02}=5.$ The other parameters are as in figure 2. Dimensionless units.\\
~\\
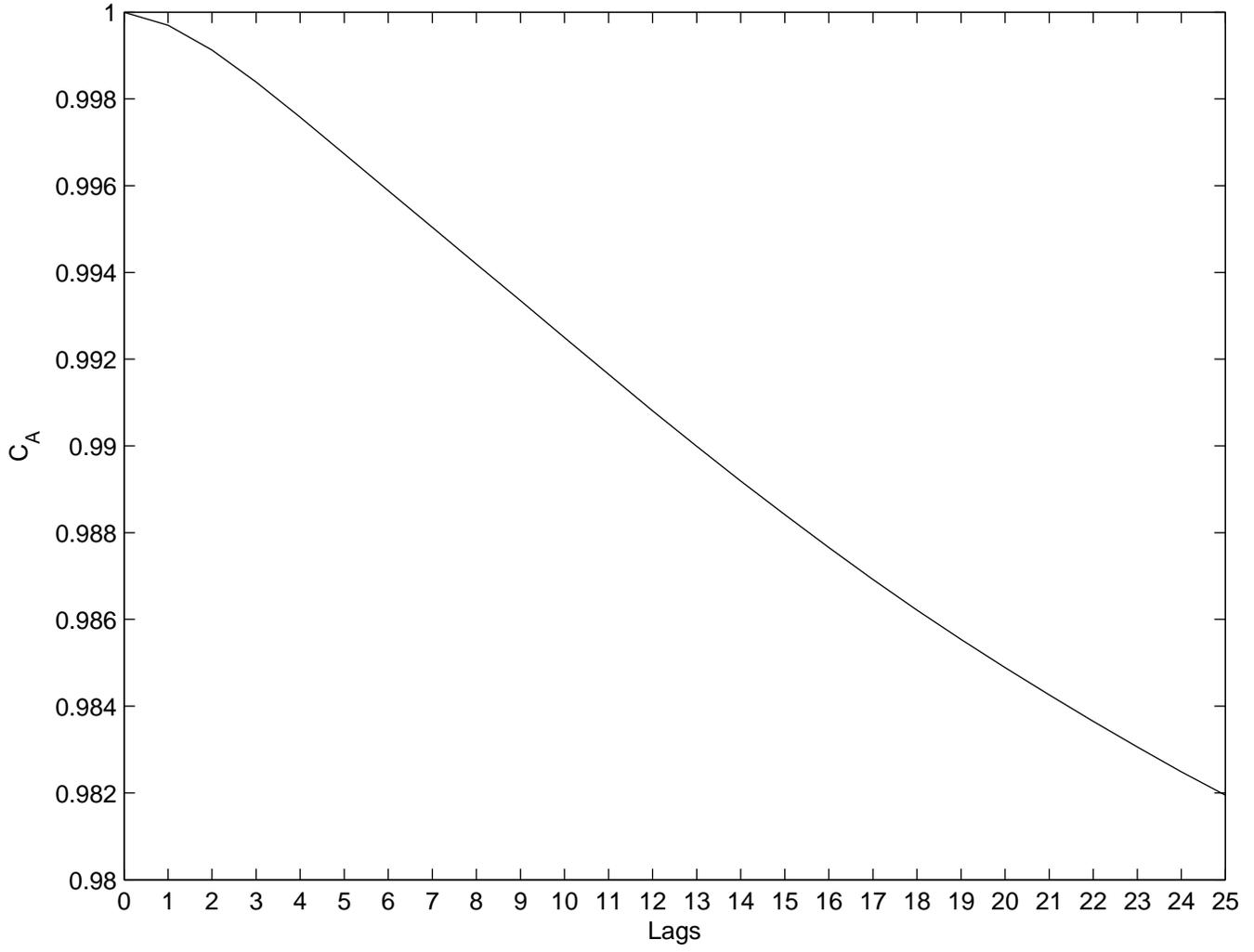
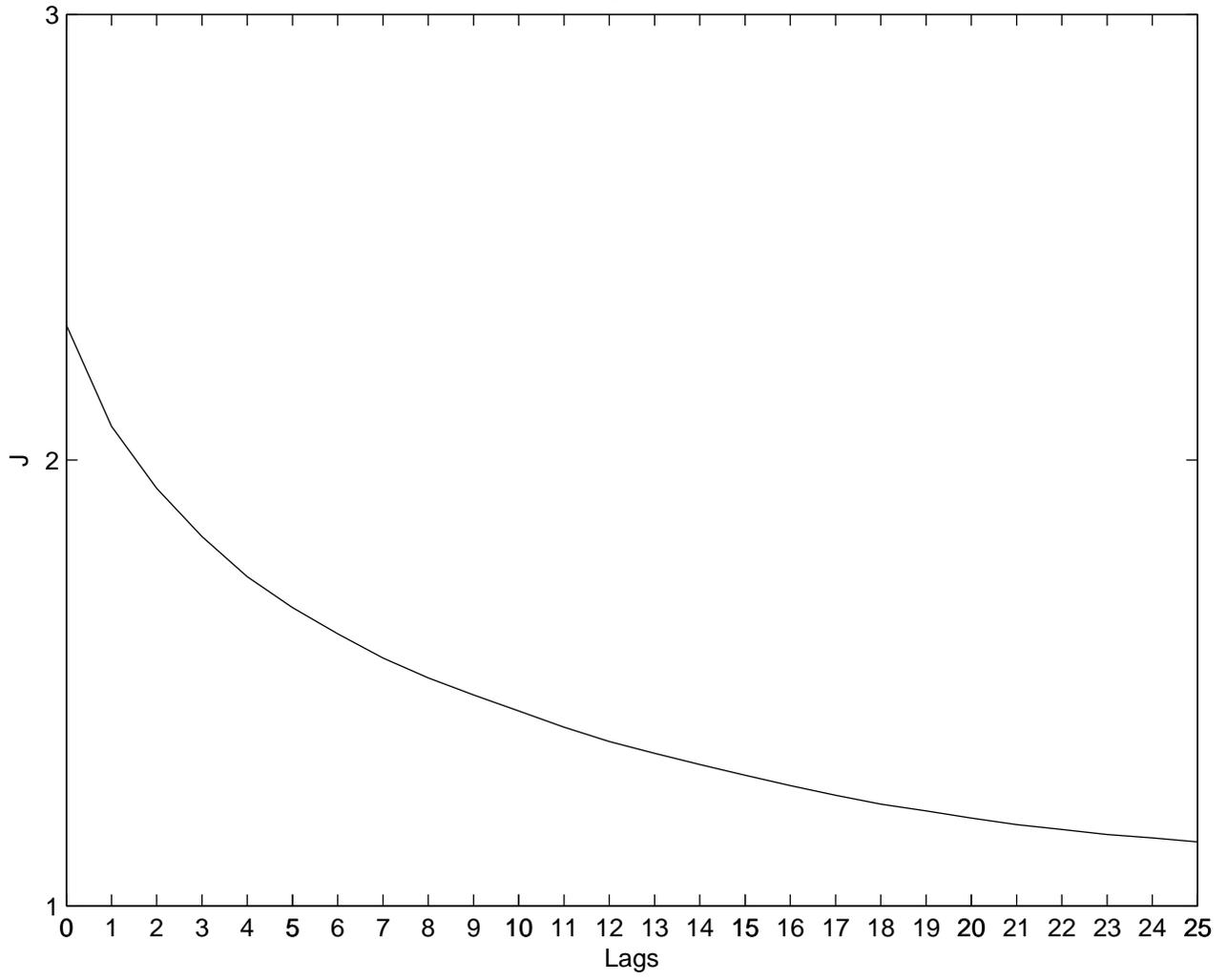
\noindent FIG.5. The autocorrelation coefficient $C_{A}$ (5(a)) and mutual information $J$ (5(b)) of the laser output for combined sinusoidal and chaotic time delays, Eq.(1) for 
$\tau_{1}(t)=3 + 0.03x_{1}(t)\sin(0.006t)$ and $\tau_{2}(t)=5 + 0.03x_{1}(t)\sin(0.006t).$ The other parameters are as in figure 2. Dimensionless units.\\
~\\
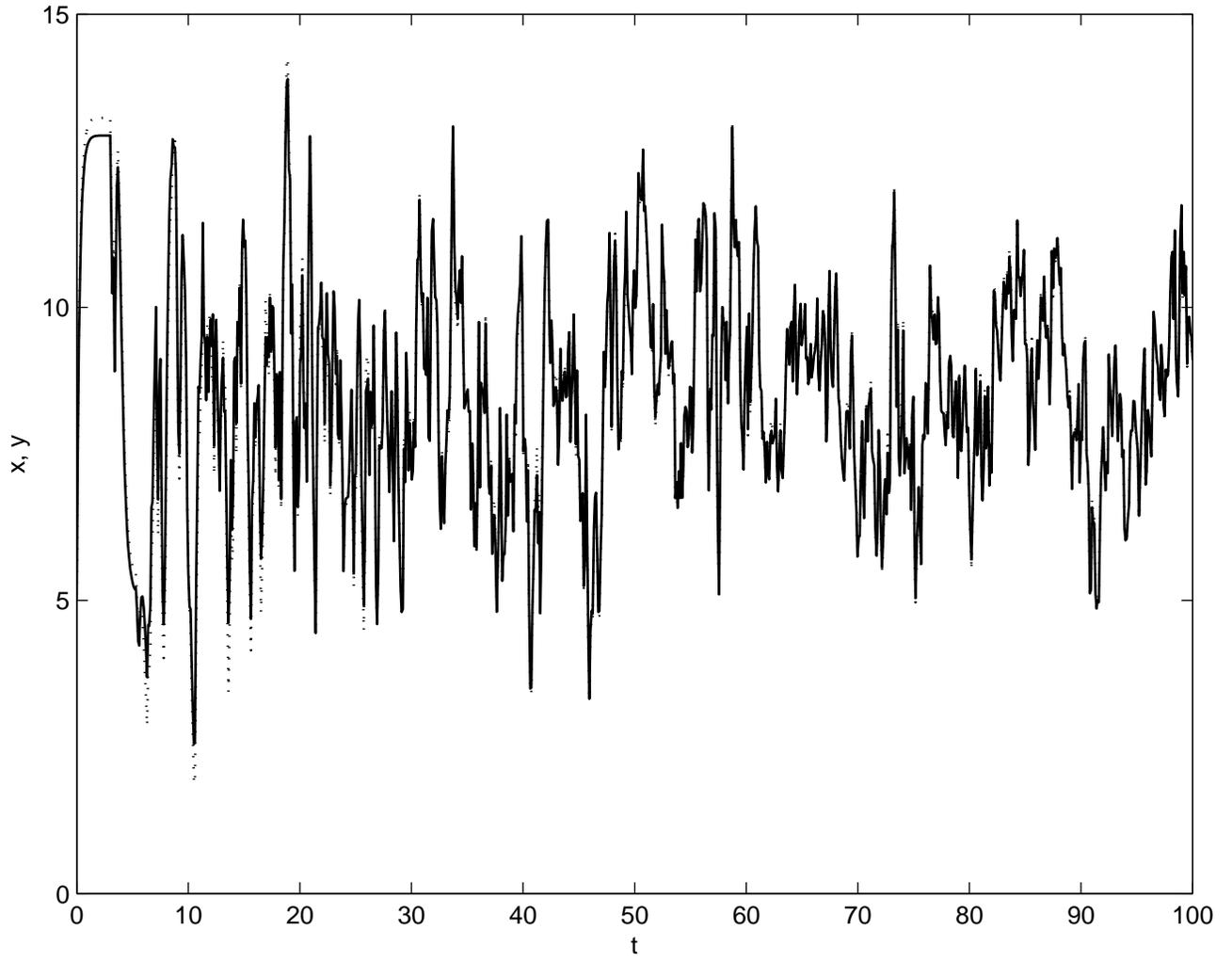
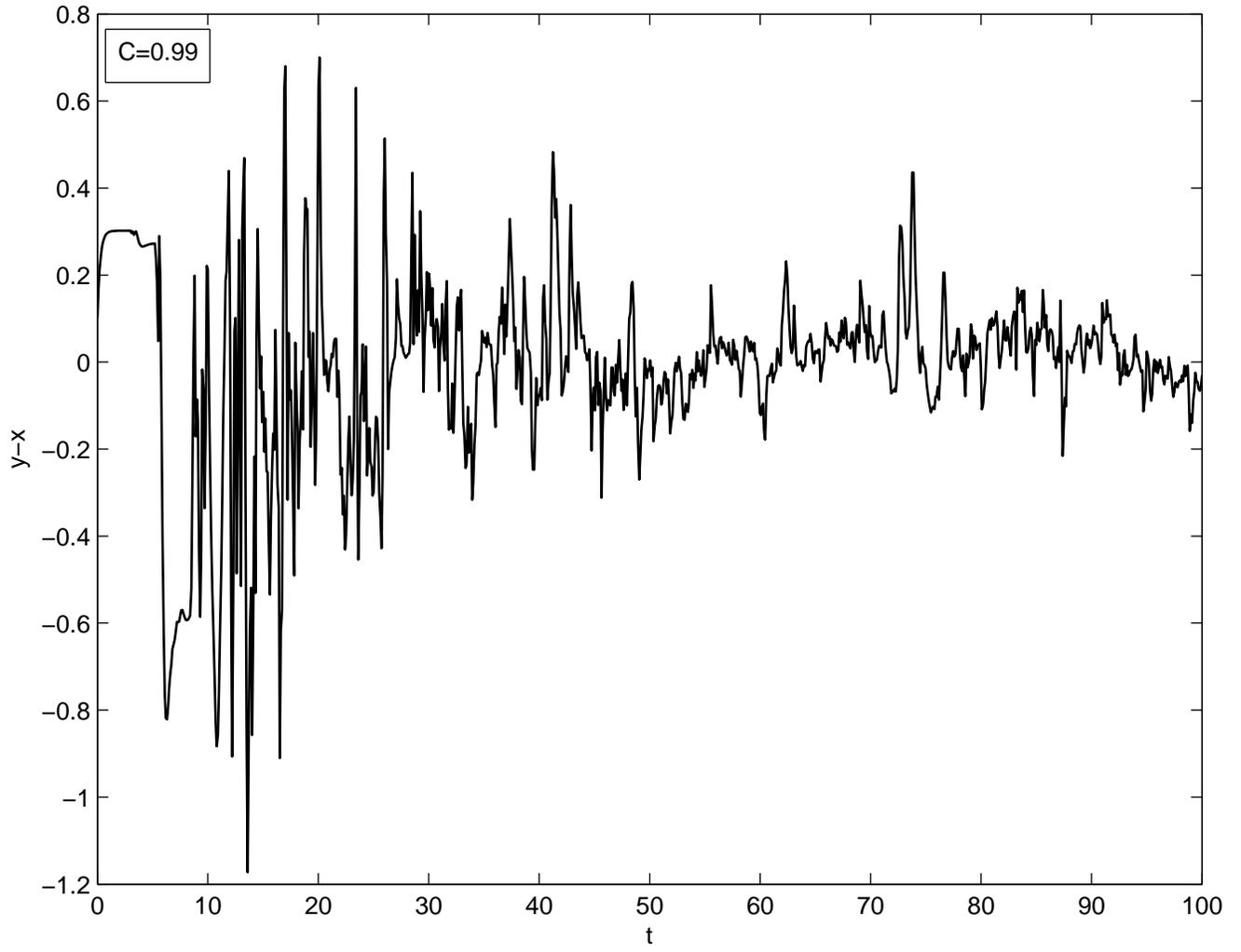
\noindent FIG.6. Numerical simulation of unidirectionally coupled variable time delay lasers, Eqs.(1-2) for $\alpha_{1}=\alpha_{2}=4,\Phi=\pi/4,m_{1}=12,m_{2}=15,m_{3}=0.5,m_{4}=15, K =11.5$ and $\tau_{1}(t)=3 + 0.03x_{1}(t)\sin(0.006t),\tau_{2}(t)=5 + 0.03x_{1}(t)\sin(0.006t),
\tau_{3}(t)=3 + 0.03x_{1}(t)\sin(0.006t)).$  Complete synchronization: 6(a): Time series of the transmitter laser wavelength (x-solid line) and receiver laser wavelength (y-dotted line);
6(b):synchronization error $y-x$ dynamics. C is the cross-correlation coefficient between the wavelengths of the transmitter and receiver lasers. Dimensionless units.\\
~\\
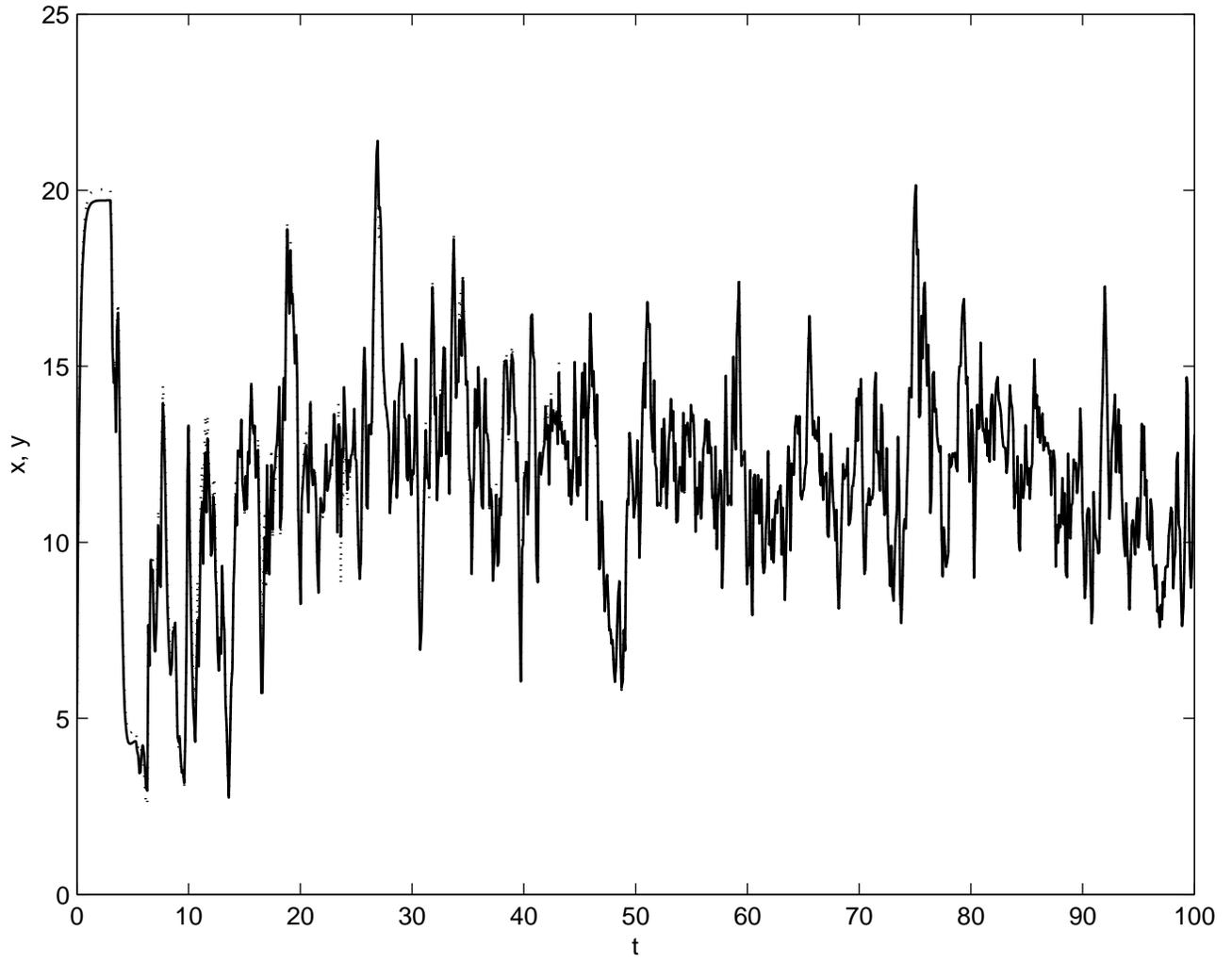
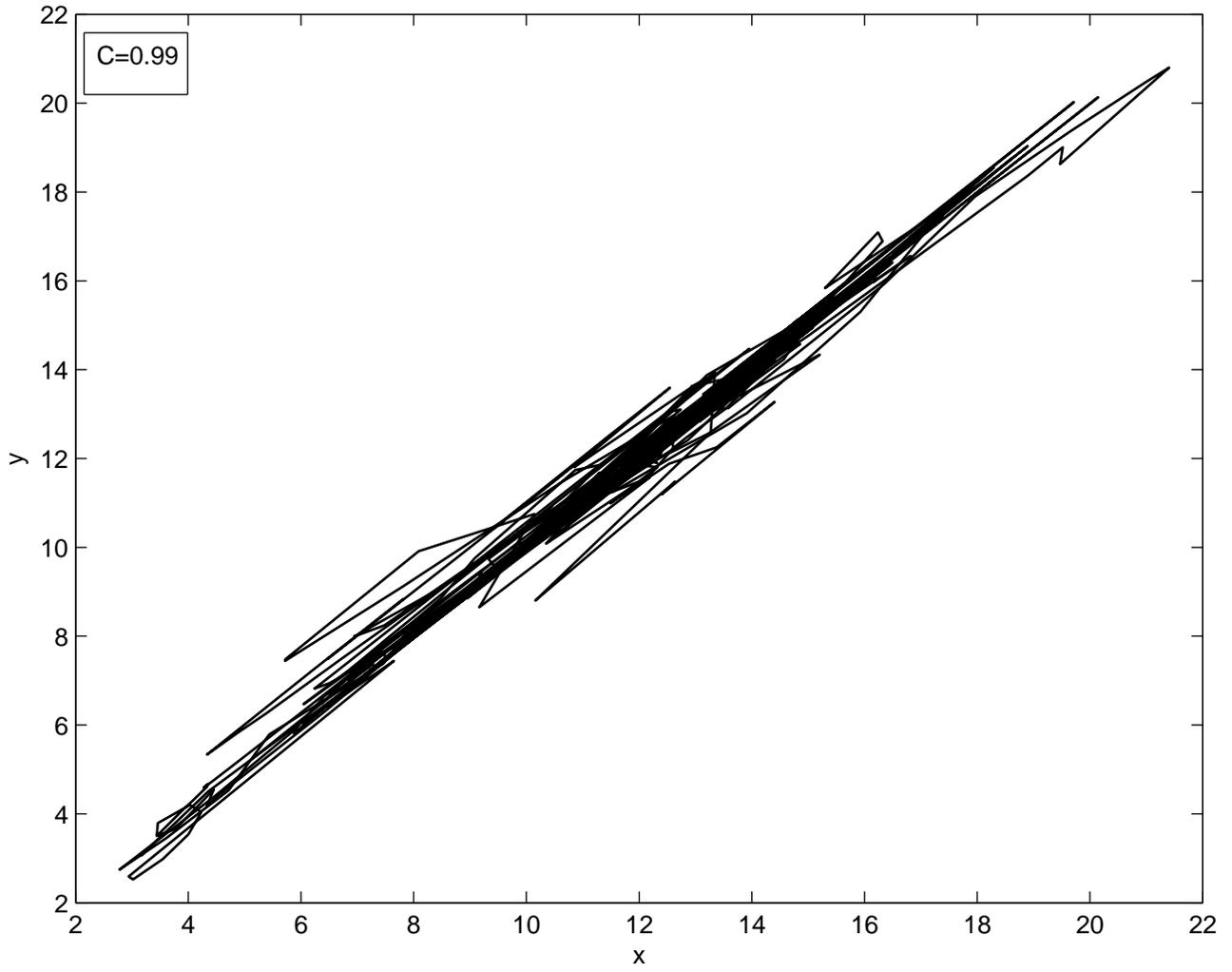
\noindent FIG.7. Numerical simulation of bidirectionally coupled variable time delays lasers, Eqs.(1-2) for $m_{1}=m_{3}=12, m_{2}=m_{4}=15,K=12.5.$ The other parameters are as in figure 6. Notice that for the mutually coupled variable time delays lasers the coupling term in Eq.(2) should be added to the right-hand side of Eq.(1).
Complete synchronization: 7(a):Time series of the $x$ laser wavelength (solid line) and $y$ laser wavelength (dotted line);7(b):y versus x.C is the cross-correlation coefficient between the wavelengths of the transmitter and receiver lasers. Dimensionless units.\\
~\\
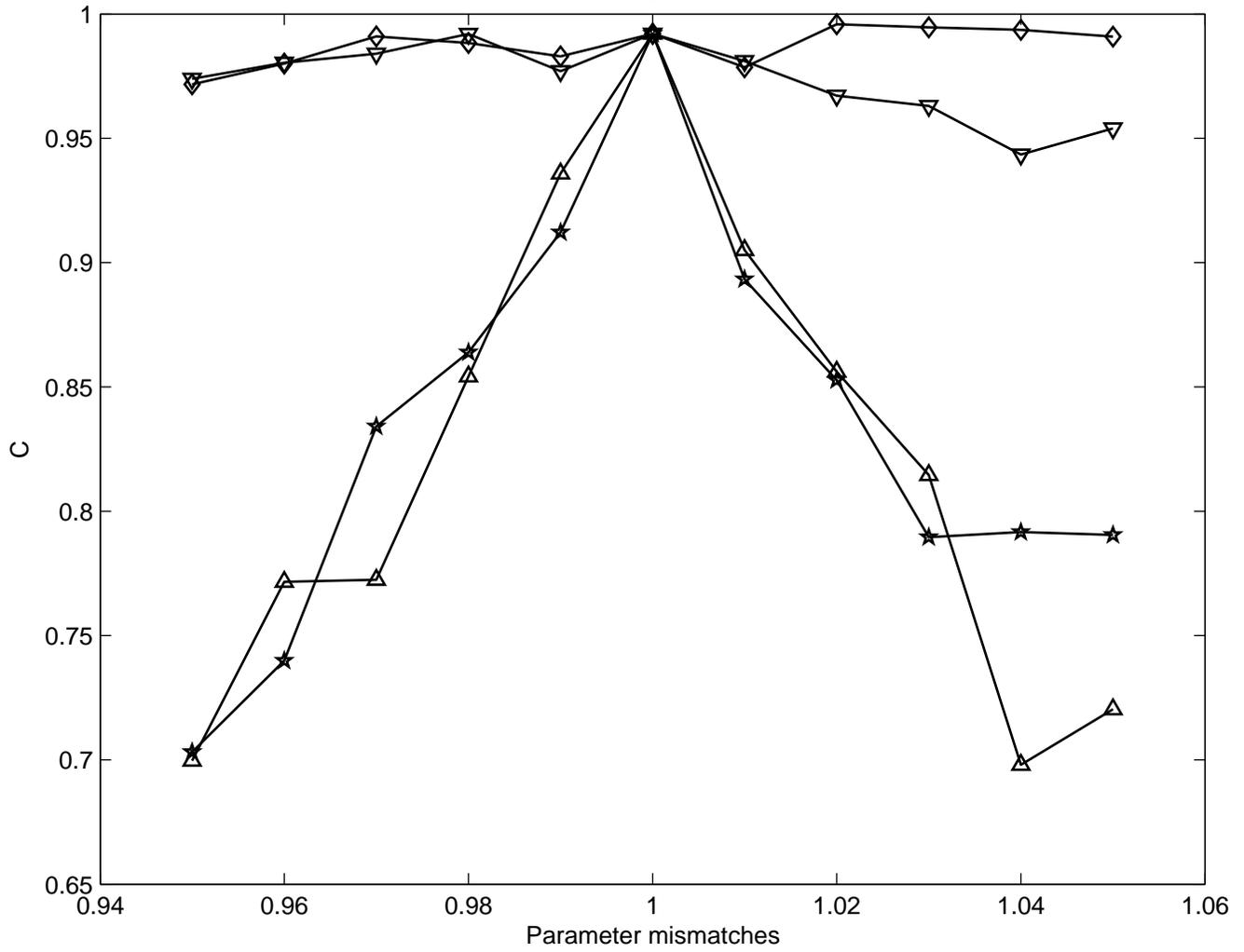
\noindent FIG.8. Numerical simulations of Eqs.(1-2) for unidirectionally coupled lasers. Dependence of the cross-correlation coefficient $C$ between the unidirectionally coupled lasers on the ratio  
$\frac{m_{2}}{m_{4}}$
of the feedback strength of the transmitter $m_{2}$ to the feedback strength of the receiver $m_{4}$($\diamondsuit$), on the ratio $\frac{\alpha_{1}}{\alpha_{2}}$
 of the relaxation coefficient of the transmitter laser $\alpha_{1}$ to the relaxation coefficient of the receiver laser $\alpha_{2}$($\nabla$), on the ratio $\frac{\omega_{1}}{\omega_{3}}$
 of the frequency of the 
feedback modulation of the tranmitter $\omega_{1}$ to the frequency of modulation of the injection term $\omega_{3}$($\triangle$), and on the ratio $\frac{\tau_{01}}{\tau_{03}}$
  of the fixed time delay of the transmitter laser $\tau_{01}$ to the fixed time delay of the injection term $\tau_{03}$($\star$).\\
~\\
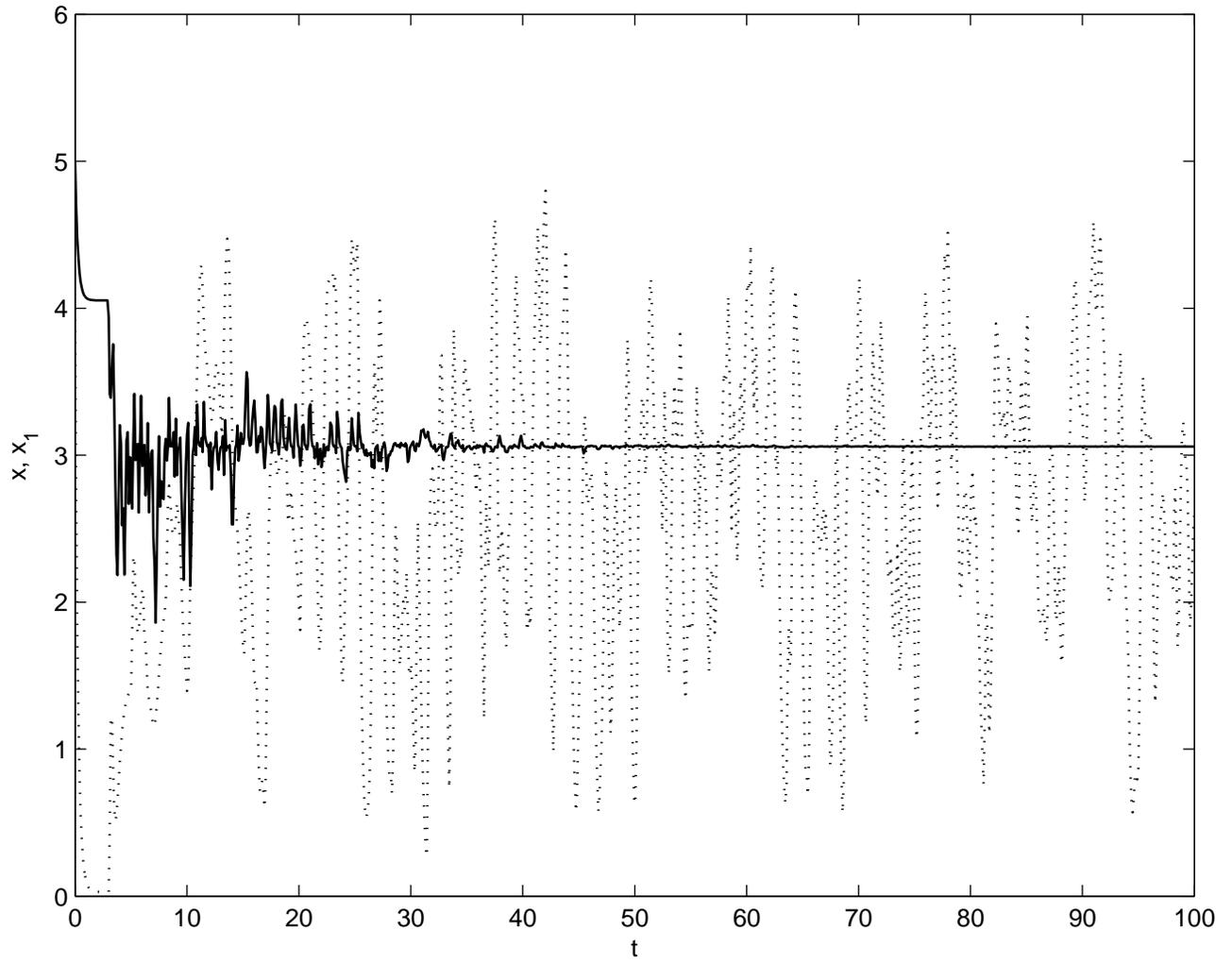
\noindent FIG.9.Numerical simulation of the variable time delay lasers, Eqs.(1) for $m_{1}=12, m_{2}=15, 
\tau_{1}(t)=3 + 0.03x_{1}(t)\sin(10t)$ and $\tau_{2}(t)=5 + 0.03x_{1}(t)\sin(10t).$
Chaos control via variable time delays:Time series of the $x$ laser wavelength with variable time delays (x(t)-solid line) 
and with fixed time delays $\tau_{01}=3, \tau_{02}=5$ ($x_{1}$(t)-dotted line).\\

\end{document}